\documentclass[reprint,prc,aps,amsmath,amssymb,showpacs,showkeys,groupedaddress,floatfix]{revtex4-2}

\usepackage{epsfig}
\usepackage{dcolumn}
\usepackage{bm}
\usepackage{graphics}
\usepackage{graphicx}
\usepackage{amssymb}
\usepackage{amsmath}
\usepackage{epstopdf}
\usepackage{subfigure}
\usepackage{float}
\newcommand{\upperRomannumeral}[1]{\uppercase\expandafter{\romannumeral#1}}

\DeclareUnicodeCharacter{2212}{-}
\begin{document}
\title{Artificial Intelligence Supported Shell-Model Calculations for Light Sn Isotopes}
\author{Serkan Akkoyun$^{1}$}
\email{ E-mail:sakkoyun@cumhuriyet.edu.tr}
\author{Abderrahmane Yakhelef$^{2,3}$}
\affiliation{$^{1}$Department of Physics, Faculty of Science, Sivas Cumhuriyet University, 58140, Sivas, Turkey}
\affiliation{$^{2}$Department of Physics, Ferhat Abbas Setif 1 University, Algeria}
\affiliation{$^{3}$PRIMALab laboratory, El-Hadj Lakhedar Batna1 University, Algeria}

\begin{abstract}
The region around the doubly magic nuclide $^{100}Sn$ is very interesting for nuclear physics studies in terms of structure, reaction and nuclear astrophysics. The main ingredients in nuclear structure studies using the shell model are the single-particle energies and the two-body matrix elements. To obtain the former, experimental data of $^{101}Sn$ isotope spectrum are necessary. Since there is not enough experimental data, different approaches are used in the literature to obtain single-particle energies. In sn100pn interaction, the hole excitation spectrum was used in $^{131}Sn$ to determine neutron single-particle energies. The other approach is the use of the lightest isotope, $^{107}Sn$, which figures the model space orbitals. In this study, we estimated the spectrum of the $^{101}Sn$ isotope by artificial neural network method in order to obtain neutron single-particle energies. After the training was carried out by using the experimental spectra of the nuclei around $^{100}Sn$ isotope, the $^{101}Sn$ spectrum was obtained. Subsequently, neutron SPEs of the model space orbitals are defined. Shell model calculations for $^{102-108}Sn$ isotopes are carried out and results are compared to the experimental data and results obtained using the widely used interaction in the region, sn100pn. According to the results, it is seen that the Sn isotope spectra obtained with the new SPE values are more compatible with the experimental data.
\end{abstract}


\maketitle
\section{Introduction}

The region around the $^{100}Sn$ isotope is one of the unique regions in the investigations of neutron deficient nuclei far from the beta stability line and close to the proton drip-line line. The $^{100}Sn$ isotope, known as the heaviest double magic nucleus close to the N=Z line, is very interesting in terms of many factors such as shell evolution, change of collective properties, band termination, and magnetic rotation. With the use of high-power radiation detectors and radioactive ion beams, experimental studies in this region have also begun to be carried out. Experimental data in the $^{100}Sn$ isotope region has begun to enrich through recent experimental studies. As the experimental excited energy information of the nuclei in this region gets richer, the results of the studies to be carried out with theoretical models such as the nuclear shell model approach to real experimental values. This allows more accurate approaches to be performed in nuclear structure studies and enables models to be examined with higher accuracy \cite{holt, insolia,sun}.
 
However, until today, in the theoretical studies carried out using the nuclear shell model in the $^{100}Sn$ region, there is not enough experimental data in the literature to obtain single-particle energy values with great accuracy. In the theoretical studies to be carried out on lighter Sn isotopes, since there is not enough experimental data for the $^{101}Sn$ isotope, different approaches have been used in the calculations to obtain neutron single-particle energy values. Yakhelef and Bouldjedri \cite{yakhelef} obtained the neutron single-particle energies by using the spectrum of the closest odd Sn ($^{107}Sn$) isotope whose experimental data are available. Brown et al. \cite{brown} obtained neutron single-particle energies by using the experimental spectrum of the $^{131}Sn$ isotope. The interaction called sn100pn in that study was derived from a realistic interaction developed starting from the G-matrix derived from the CD Bonn nucleon-nucleon interaction. A set of interactions named snet by Hosaka et al. \cite{hosaka} is derived from a bare G-matrix based on the renormalized Paris potential for N=82 nuclei. The single-particle energies of these sn100pn and snet interactions are widely used in the literature for the shell-model calculations performed on the A=100 region. Trivedi et al. \cite{trivedi} modified these commonly used neutron single-particle energies with the value of the $7/2^{+}$ level, which is the only experimental data for the 101Sn isotope. Leander et al. \cite{leander} theoretically obtained neutron single-particle energies, based on Hartree-Fock with Skyrme III interaction, folded Yukawa potential and Wood-Saxon single-particle potentials, separately. Engeland et al. \cite{engeland} modified the previous work mentioned by using experimental observations. Andreozzi et al. \cite{andreozzi} resorted to the analysis of low energy spectra of isotopes with A<111, since there is not enough information in the literature regarding the spectrum of the $^{101}Sn$ isotope. Sandulescu et al. \cite{sandulescu} obtained neutron single-particle energies by fitting known one quasi-particle excitation at $^{111}Sn$. Grawe et al. \cite{grawe} and Schubart et al. \cite{schubart} used $^{88}Sr$ or $^{90}Zr$ isotopes to theoretically obtain neutron single-particle energies for this region.

The lack of experimental data on the $^{101}Sn$ isotope in the literature is our main motivation for this study. We aimed to address the problem with a completely different approach from those previous studies. For this purpose, we considered artificial neural networks (ANN) to be used as a tool. Our goal is to obtain the excited energy spectrum of the $^{101}Sn$ isotope, which allows us to extract neutron single-particle energies to be used in the shell-model calculations. So what we need to do is obtain the values of the $5/2^{+}$, $7/2^{+}$, $1/2^{+}$, $3/2^{+}$ and $11/2^{-}$ excited energy levels of the $^{101}Sn$ isotope with the highest possible accuracy. In cases where this cannot be achieved experimentally, the ANN method \cite{haykin} appears to be very good. We thought that we would be able to obtain these energy level values of the $^{101}Sn$ isotope with this method, as was the case with previous studies on the static structural properties of nuclei. In these previous studies we mentioned, we were able to obtain the ground-state energies of all nuclei in the nuclear chart \cite{bayram}, the first excited $2^{+}$ energy values of even-even nuclei \cite{akkoyun1} and the radii of all nuclei \cite{akkoyun2} with great accuracy. These results showed us that the ANN method can be used as an alternative tool in investigating the structure of atomic nuclei. Therefore, this indicates that the excited energy values of the $^{101}Sn$ isotope can also be obtained with great accuracy. We confirmed this expectation by using the neutron single-particle energies we obtained from this present study in the shell-model calculations. We compared our results of the shell-model calculations performed on light isotopes of $^{102-108}Sn$ with those performed with neutron single-particle energies widely discussed in the literature. As can be seen, the results from these neutron single-particle energies obtained with the support of ANN method are closer to experimental values compared to those in the literature. Thus, for the first time in the literature, we think that we have obtained neutron single-particle values with accuracy using such a different approach.

\section{Artificial Neural Networks}

Artificial neural network (ANN) \cite{haykin} is one of the most powerful methods preferred in case of missing unknown data. It is a mathematical tool that mimics brain functionality \cite{hornik}. ANN has processing elements (neurons) in different layers which are input, hidden and output layers. Neurons in the input layer receive data and transmit it to hidden layer neurons and then to output layer neurons. Due to layered structure and forward data flow, this type of ANN is called a layered feed-forward ANN that we have considered in the present study. The neurons in a layer are connected to each neuron in the next layer by adjustable synaptic weights. The main purpose of the ANN method is to determine the appropriate values for the weights by training the given sample data. The numbers of neurons in the input and output layers depend on the problem. Besides, there is no rule to determine the number of hidden layers and their neurons. The hidden layer neuron number is determined after several trials that give the best results for the problem. 

In this study, we used a structure with 4 (or 5) input neurons, 6 hidden layer neurons and 1 output neuron in the ANN method to estimate some excited levels of the $^{101}Sn$ isotope. These levels are the $1/2^{+}$, $3/2^{+}$ and $11/2^{-}$ levels needed for the shell-model calculations. We handled the experimental data of these levels of isotopes in the sdgh model space around A=100 as data in the ANN method. We took 3 different approaches in our calculations. The experimental data we used in the first approximation were only those belonging to odd Sn isotopes in the mass range of 101-131 (set-1). In this approach, 56 of all data were used for the training and 14 for the test. In another approach, we included $^{101-131}Sn$, $^{105-131}Te$, $^{113-133}Xe$ and $^{119-135}Ba$ isotopes in the dataset, which are in this neighborhood with odd neutron and even proton numbers and have experimental data in the literature (set-2). In this approach, 172 of all data were used for training and 42 for the test. In the last approach, we used the available experimental data of all odd and even proton number isotopes $^{101-131}Sn$, $^{105-131}Sb$, $^{105-131}Te$, $^{109-135}I$, $^{113-133}Xe$, $^{113-137}Cs$ and $^{119-135}Ba$ in the neighbourhood of A=100 (set-3). In this approach, 293 of all data were used for training and 72 for testing. In all these approaches, the number of values to be produced by ANN was 3. These are the energy values of the $1/2^{+}$, $3/2^{+}$ and $11/2^{-}$ levels belonging to the $^{101}Sn$ isotope. In all three approaches, experimental data of the existing 5/2+, $7/2^{+}$, $1/2^{+}$, $3/2^{+}$ and $11/2^{-}$ levels of the isotopes included in the data set were used. ANN inputs are mass number (A), neutron number (N), proton number (Z) (except for the first approximation because this parameter is not discriminatory there), orbital angular momentum number (L), and total spin (J). The output of ANN is the aforementioned excited level energy. After many trials, a structure with 6 hidden neurons that gives the best result was considered and tangent hyperbolic function was used as an activation function.

\begin{figure}
\includegraphics[width=8cm]{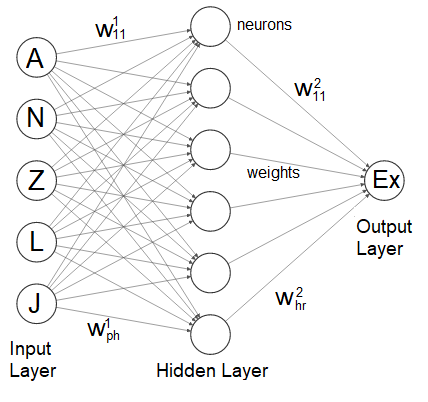}
\caption{\label{Fig1} ANN structure with 5-6-1 topology for the prediction of $^{101}Sn$ excited energy state energies}
\end{figure}

Here, we give the minimum ANN fundamentals. For ANN with a single hidden layer, the desired output vector $\vec{y}$ (Ex. in Fig.1) is approximated by a network multi-output vector $\vec{f}$. The multi-output vector is defined by Eq. (1) as given below.

\begin{eqnarray}\label{ann1}
\vec{f}:R^{p}\rightarrow R^{r}:\vec{f}_{k}(\vec{x)}=\sum_{j=1}^{h_{1}} \beta_{j}G(A_{j}(\vec{x})), \nonumber\\
\vec{x}\in R^{p},\beta_{j}\in R, A_{j}\in A^{p}, and k=1,...,r
\end{eqnarray}
where $A^{p}$ is the set of all functions of $R^{p}\rightarrow R^{r}$ defined by  $A(\vec{x})=\vec{w}.\vec{x}+b$, $\vec{w}$ is weight vector from input layer to hidden layer, $\vec{x}$ is the input vector of ANN (4 or 5 inputs in Fig.\ref{Fig1}), $b$ is the bias weight, and $p(r)$ number corresponds to each input (output) variable. In this study, we have used four or five input layer neurons ($p=4 or 5$), one output layer neuron ($r=1$), and six hidden layer neurons ($h=6$) (see Fig. 1). The total number of adjustable weights ($\sum W$) are calculated by Eq. (2) as 30 or 36 for four or five input neurons, respectively.

\begin{eqnarray}\label{ann2}
\sum W=p.h+h.r=h.(p+r)
\end{eqnarray}

In Fig. 1, the weight matrices $w^{1}$  and $w^{2}$  correspond to weight vectors defined in $A(\vec{x})$ and $\vec\beta$ in Eq.(1). However, as seen in Fig. 1 and Eq. (1), the correspondences $w^{1}\rightarrow A(\vec{x})$ and $w^{2}\rightarrow \vec\beta$  are valid only for the ANN structure with a single hidden layer. For the ANN with more than one hidden layer, both Eq.(1) and the correspondences must accordingly be changed. The activation function for hidden neurons $G:R\rightarrow R$ in Eq.(3) can be theoretically any well-behaved non-linear function. Commonly, $G$ is chosen as a non-linear sigmoid type function defined by Eq.(3). The type of activation function $G$ in Eq. (2) is hyperbolic tangent for the hidden layer in the present work.

\begin{eqnarray}\label{ann3}
G:R \rightarrow [0,1] \: or \: [-1,1], non-decreasing, \nonumber\\ lim_{\lambda\rightarrow\infty} G(\lambda)=1, lim_{\lambda\rightarrow-\infty} G(\lambda)=0 \: or -1
\end{eqnarray}

The method is also a perfect tool for non-linear function approximations. It is composed of two main stages which are training and test. The entire data belonging to the problem is divided into two separate sets for these stages. In the training stage, the first part of data is given to the ANN, including both input and desired output values. The Levenberg-Marquardt algorithm was used for the training of the ANN. The weights are modified using the sample data in this stage. The method generates its own outputs as close as possible to the desired output values. Comparisons between the desired output and the ANN output are made by root mean square error (RMSE) function given by Eq.(4)

\begin{eqnarray}\label{ann4}
RMSE=\sqrt{\sum_{i=1}^{N} (\dfrac{(y_{i}-f_{i})^{2}}{N})}
\end{eqnarray}
where $N$ is the total number of the data in the stage, $y_{i}$ is the desired output, and $f_{i}$ is the ANN output. After an acceptable error level between the ANN outputs and the desired outputs, the training stage is terminated. This means that the ANN is appropriately constructed for solving the problem with the modified final weights. However, it is still early to decide whether the constructed ANN is convenient for the estimation of similar another set of data. The generalization ability of the ANN must be tested using the second set of the data that is never seen by the constructed ANN in the training stage. If the generated outputs in the test stage by using final weights are still close to the desired outputs, it can be confidently concluded that the ANN is useful for the solution of the problem. Namely, every other data related to the existing data set might be predicted by the constructed ANN. For further details of ANN, the authors refer to read the reference \cite{haykin}.

\section{Sheel Model Calculations}

The nuclear shell model is one of the most suitable tools to describe the low-energy structure of atomic nuclei \cite{mayer,talmi,caurier1}. In this model, nucleons are assumed to move independently in a central potential well. Calculating nuclear energy levels is a very difficult task. The main reason for the difficulty is that the nature of the interaction between free protons and neutrons, in other words, strong nuclear interaction, is not well known. If we consider a nucleus with several valence nucleons outside of closed shell, the energies of the levels can be divided into three parts. First, the binding energy of closed shells ($^{100}Sn$ in this study), the second is the sum of single nucleon energies, including the kinetic energies of the valence nucleons and their interactions with the nucleons of the core. The third is the interaction of valence nucleons with each other. Of these, it is the most difficult to calculate the binding energies of closed shells. The easiest to calculate is the interaction between valence nucleons. 

The nuclear shell model is the paradigm of choice for the understanding of the nuclear structure. Its main idea consists of considering the nucleus as a quantum system constituted of  protons and neutrons moving freely in a self-generated mean-field. For the nucleus with A nucleon, the Hamiltonian can be written as in Eq.(5).

\begin{equation}\label{hamiltonian1}
H=\sum\limits_{i}^A T_{i}+\frac{1}{2} \sum\limits_{ij}^A V_{ij}
\end{equation}

Here $T_{i}$ is the kinetic energy of each nucleon, and $V_{ij}$ is the interaction potential between nucleons. But since the interaction between nucleons is not clearly defined, assuming that each nucleon moves at an average potential formed by the others, the Hamiltonian can be arranged as given in Eq.(6) and (7) by adding and subtracting a sum of single-particle potential energy, $U_{i}$.

\begin{equation}\label{hamiltonian2}
H=\sum\limits_{i}^A [T_{i}+U_{i}]+[\frac{1}{2} \sum\limits_{ij}^A V_{ij}-\sum\limits_{i}^A U_{i}]
\end{equation}

\begin{equation}\label{hamiltonian2}
H=H_{0}+H_{res}
\end{equation}
where $H_{0}$ describes the motion of the A nucleons, independent of each other, in the same average field. $H_{res}$ is the residual potential presumed to be much smaller in strength than the total potential. In this mean-field approximation, the strongly interacting A fermions system is converted to a system of A non-interacting fermions where each nucleon can be viewed as moving in an external potential $U$ created by the remaining A-1 neighbours. An approximation of this potential is the spherically symmetric potential of Wood-Saxon in which the radius and diffuseness can be adjusted for each nucleus or an entire region of nuclei. The harmonic oscillator potential is an approximation to the Wood-Saxon potential. Yukawa potential is also used as an approximation to the mean-field which gives good results. A strong spin-orbit term is included in order to obtain the spectrum of single-particle orbitals allowing the obtaining of closed shells corresponding to magic numbers \cite{mayer2,mayer3}. Within this approximation, the nucleus is considered as an inert core made up of shells filled up with neutrons and protons plus a certain number of valence nucleons. This extreme single-particle shell model, supplemented by empirical coupling rules, is sufficient for a good description of various nuclear properties, like the angular momentum and parity of the ground-states of odd-mass nuclei. However, a description for nuclei with two or more valence nucleons needs an explicit inclusion of the residual two-body interaction between valence nucleons to remove the degeneracy of the states belonging to the same configuration. If these valence nucleons are in a single orbit, it is sufficient to know only the matrix elements of the effective interaction between nucleons in that orbit. If the valence nucleons are distributed over several orbitals, differences between single nucleon energies (single-particle energies) are also needed, which can usually be taken from experimental data if available. Because of the difficulty of not knowing the individual interactions between nucleons, instead of these interactions, an average potential (mean-field approximation) generated by other nucleons is involved. Thus, the problem dealt with in the nuclear shell model, the many-body problem that takes into account all nucleons in the nucleus, is reduced to a few-body problem that only takes into account valence nucleons.

Single-particle energies can be determined by choosing a central potential such as the harmonic oscillator, Wood-Saxon or Yukawa type. The two-body matrix elements (tbme) belonging to the interaction ($H_{res}$) represent the interactions between nucleons. In this present study, we modified single-particle energies (spe) belonging to a set containing spe and tbme values, which are widely used in the literature. We made this modification via ANN, which we mentioned in the previous section. We used the excited energy level values of the $^{101}Sn$ isotope produced in ANN to obtain the neutron spe values. Thus, we have worked on the improvement of the $H_{0}$ term in the expression given by Eq.(7).

In any shell-model calculations, one has to start by defining a model space which is a set of active single-particle orbits outside the inert core. The basic inputs are the single-particle energies (SPE) of the stats of the chosen model space, and two body matrix elements (TBME). For the former, SPE is explicitly calculated using the mean-field models or defined empirically from the available experimental data of nuclei in the direct vicinity of the doubly magic nuclei. For the latter, the TBME is specified in terms of matrix elements of the residual interaction $H_{res}$, $\langle j_{1}j_{2}J|H_{res}|j_{3}j_{4}J\rangle$, for all possible combinations of ji orbitals in the model space. J is the total two particles' angular momentum. The final step in carrying out shell model calculations is to diagonalize the model space-effective interaction.

In this present study, we modified single-particle energies (spe) belonging to a set containing spe and tbme values, which are widely used in the literature. We made this modification via ANN, which we mentioned in the previous section. We used the excited energy level values of the $^{101}Sn$ isotope produced in ANN to obtain the neutron spe values. Thus, we have worked on the improvement of the $H_{0}$ term in the expression given by Eq.(7). For the shell-model calculations performed in matrix formalism for many-particle systems, as the size of the model space and the number of nucleons increase, the dimensions of the Hamiltonian matrix increase to very high orders ($10^{10}$). To obtain eigenvalues, matrices are diagonalized using appropriate algorithms such as Lanczos and the solution is reached. For this purpose, there are many computer codes developed to perform nuclear shell model calculations in the literature. Examples of these are Bigstick \cite{jhonson}, Kshell \cite{shimizu}, Oxbash \cite{oxbash}, Antoine \cite{caurier2} and Nushell \cite{brown2}. In the calculations performed in this study, the Kshell code was used. This code, which runs on Linux operating system, allows performing nuclear shell model calculations with M-scheme representation using the Lanczos method. Energy levels of nuclei, spin and isospins, magnetic and quadrupole moments, $B(E2)$ and $B(M1)$ transition probabilities between levels and single-particle spectroscopic factors can be calculated up to $10^{10}$ size with the code.

\section{Result and Discussions}

\subsection{Obtaining neutron spe values by ANN}

In the study, firstly, some excited energy states of $^{101}Sn$ isotope required to obtain neutron single-particle energies were obtained by using ANN method. The energy states of the $^{101}Sn$ isotope obtained with ANN are shown in Table 1 for different data sets mentioned in section 2. To obtain neutron single-particle energies, it is more appropriate to study neutron single-particle states in the sdgh shell, we first thought of isotopes with only valance neutrons in this shell. For this purpose, we hypothesized Sn isotopes with only neutrons as well as having no protons in the sdgh shell. We have compiled data on the first $5/2^{+}$, $7/2^{+}$, $1/2^{+}$, $3/2^{+}$ and $11/2^{-}$ energy levels of Sn isotopes in the 101-131 mass number range, whose experimental excited state energy values are available in the literature. We randomly allocated these data to use $80\%$ in the training of the ANN and $20\%$ in the test stage. In the calculations made for this set (set-1), atomic masses (Z) and neutron (N) numbers of Sn isotopes, angular momentum (L)  numbers of excited levels and total angular momentum (J) of excited levels were used as inputs of ANN. The output of the ANN is the excited state energies of the $^{101}Sn$ isotope. MSE values of training and test stages were obtained as 0.095 and 0.132 MeV, respectively. Correlation coefficients (r) were obtained as 0.99 for the training and 0.93 for the test stage. Using the experimental binding energies of $^{101}Sn$ and $^{101}Sn$ isotopes, we calculated the binding energy of the $5/2^{+}$ level as -11.0939 MeV. In addition, the energy of the $7/2^{+}$ level of $^{101}Sn$ isotope is also available in the literature. Based on these values, we calculated the neutron single-particle energies of $d_{5/2}$, $g_{7/2}$, $s_{1/2}$, $d_{3/2}$ and $h_{11/2}$ as given in Table 2. In this table, we have also given the neutron spe values in sn100pn interaction which is widely used in the shell model calculations for these isotopes.

\begin{table}[H]
\caption{\label{table1} Some Energy states of the $^{101}Sn$ isotope required for neutron single-particle energies in the region A=100.}
\begin{ruledtabular}
\begin{tabular}{cccc}
$J^{\pi}$  & $1/2^{+}$ & $3/2^{+}$ & $11/2^{-}$\\
\hline
Set-1     & $2.600$ & $1.833$ & $2.434$\\
\hline
Set-2     & $0.883$ & $1.353$ & $2.222$\\
\hline
Set-3     & $1.158$ & $1.429$ & $1.942$\\
\end{tabular}
\end{ruledtabular}
\end{table}

In the second part of the study to obtain excited levels of the $^{101}Sn$ isotope by ANN, we used the experimental energies of the first $5/2^{+}$, $7/2^{+}$, $1/2^{+}$, $3/2^{+}$ and $11/2^{-}$ levels with even protons in the sdgh shell, but also with odd neutron numbers. The purpose of ANN calculations is to find the excited levels of this isotope accurately. Increasing the number of correct or closest data used for the training of the ANN will affect the quality of the training. Therefore, in this approach where we use set-2 data, we thought of expanding our data set a little more and performing ANN training with a large data set. We obtained the first $5/2^{+}$, $7/2^{+}$, $5/2^{+}$, $3/2^{+}$ and $11/2^{-}$ energy level values of Sn, Te, Xe and Ba isotopes, whose experimental excited energy values are available in the literature. We randomly divided these data (set-2) into two parts, $80\%$ of which will be used in the training and $20\%$ in the test stage. In the calculations, atomic masses, neutron numbers, proton numbers of the isotope, angular momentum numbers and total angular momentum of excited levels were used as inputs of ANN. The output of the ANN is the excited energies of 101Sn isotope. MSE values of training and test stages were obtained as 0.101 and 0.122 MeV, respectively. Correlation coefficients (r) were found as 0.98 for the training stage and 0.96 for the testing stage. In this approach, we obtained the neutron single-particle energies of $d_{5/2}$, $g_{7/2}$, $s_{1/2}$, $d_{3/2}$ and $h_{11/2}$ as given in Table 2.

In the last part of our studies to obtain the excited levels of the $^{101}Sn$ isotope by ANN, we used the experimental energies of the first $5/2^{+}$, $7/2^{+}$, $1/2^{+}$, $3/2^{+}$ and $11/2^{-}$ levels with both even and odd proton numbers in the sdgh shell and odd neutron numbers. We have made this broadening in our data in order to increase the number of data used for training of the ANN a little more as possible. We obtained the first $5/2^{+}$, $7/2^{+}$, $5/2^{+}$, $3/2^{+}$ and $11/2^{-}$  energy level values of Sn, Sb, Te, I, Xe, Cs and Ba isotopes, whose experimental level energy values are available in the literature. We again randomly divided these data (set-3) into two parts, $80\%$ of which will be used in training and $20\%$ in the test stage. In the calculations, atomic masses, neutron numbers, proton numbers of the isotopes, angular momentum numbers and total angular momentum of the excited levels were used as inputs of ANN. The output of the ANN is the energies of the aforementioned excited levels of the 101Sn isotope. MSE values of training and test stages were obtained as 0.123 and 0.136 MeV, respectively. Correlation coefficients (r) were obtained as 0.96 for the training stage and 0.92 for the testing stage. The $d_{5/2}$, $g_{7/2}$, $s_{1/2}$, $d_{3/2}$ and $h_{11/2}$ neutron single-particle energies calculated in this approach are also given in Table 2.

\begin{table}[H]
\caption{\label{table2} Neutron single-particle energy values for shell model calculations in the region A=100.}
\begin{ruledtabular}
\begin{tabular}{cccccc}
$$  & $d_{5/2}$ & $g_{7/2}$ & $s_{1/2}$ & $d_{3/2}$ & $h_{11/2}$  \\
\hline
Set-1     & $-11.0939$ & $-10.9222$ & $-8.4939$ & $-9.2609$ & $-8.6599$\\
\hline
Set-2     & $-11.0939$ & $-10.9222$ & $-10.2109$ & $-9.7409$ & $-8.8719$\\
\hline
Set-3     & $-11.0939$ & $-10.9222$ & $-9.9359$ & $-9.6649$ & $-9.1519$\\
\hline
sn100pn     & $-10.2893$ & $-10.6089$ & $-8.6944$ & $-8.7167$ & $-8.8152$\\$$
\end{tabular}
\end{ruledtabular}
\end{table}

\begin{figure}
\includegraphics[width=10cm]{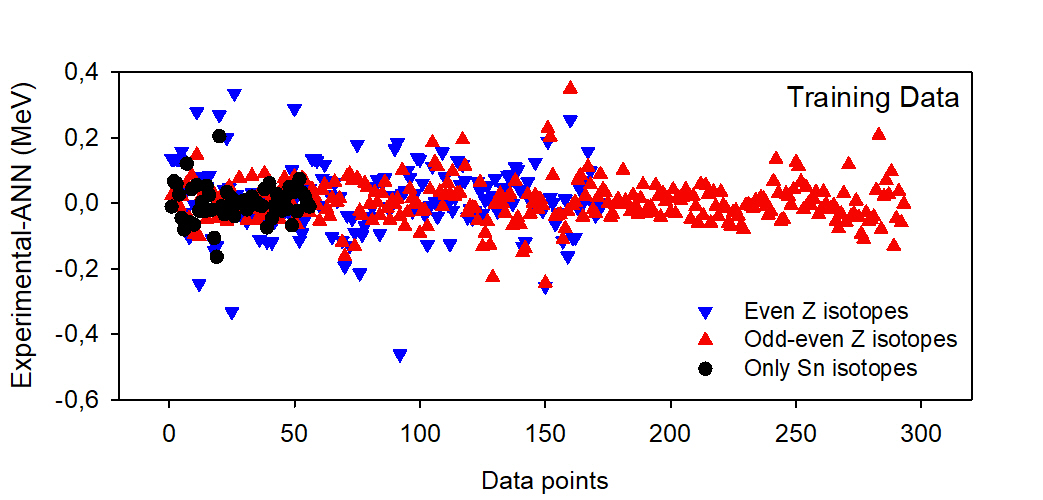}
\caption{\label{Fig2} The difference between experimental and ANN estimation energy values on the training data for the first $5/2^{+}$, $7/2^{+}$, $1/2^{+}$, $3/2^{+}$ and $11/2^{-}$ energy levels around $^{100}Sn$ region.}
\end{figure}

In Fig.\ref{Fig2}, the results of ANN calculations performed with all three data sets are presented on the training data of the ANN. The differences of ANN estimations of all three sets from the experimental data in the data sets are shown on the same graph. As can be clearly seen from the figure, these differences are concentrated around the zero line. The maximum value of the difference in the data set, where only even Z-numbered and odd N-numbered isotopes are considered, is around 0.5 keV. The number of points with a difference of more than 0.2 keV is quite small. The maximum value of the difference in the data set using both odd and even Z and odd N isotopes is approximately 0.4 keV. In addition, in the case where only the experimental data of Sn isotopes are used, the maximum deviation of ANN estimates is about 0.2 keV.

\begin{figure}
\includegraphics[width=10cm]{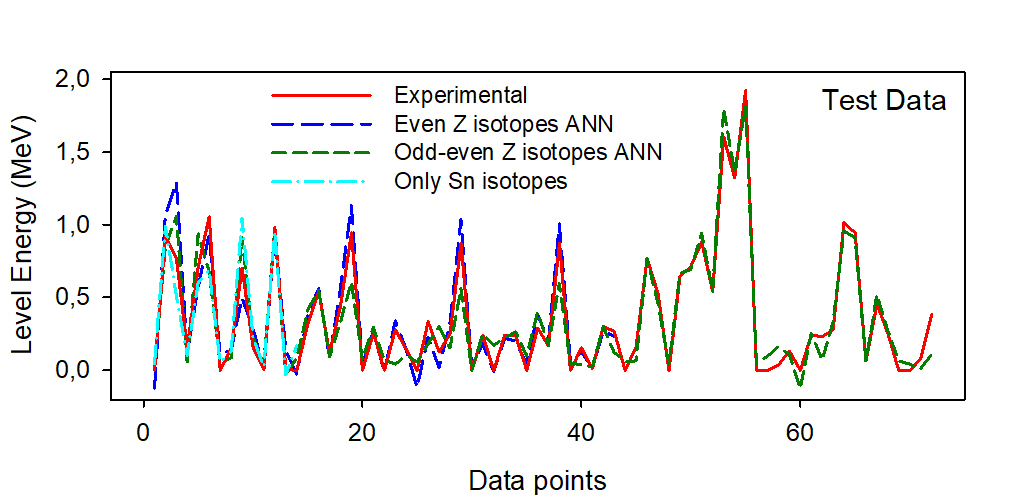}
\caption{\label{Fig3} The experimental and ANN estimation energy values on the test data for the first $5/2^{+}$, $7/2^{+}$, $1/2^{+}$, $3/2^{+}$ and $11/2^{-}$ energy levels around $^{100}Sn$ region.}
\end{figure}

The results of the test of the ANN generated are given in Fig.\ref{Fig3}. It is seen that the results of the calculations performed with all three sets are in accordance with the experimental data. This agreement indicates that the ANN method is successful in obtaining the first excited states of isotopes in the $^{100}Sn$ region. This shows that the ANN method is suitable for generating energy levels of $^{101}Sn$ isotopes, which are not available in the literature.

To once again test the accuracy of ANN predictions, we also examined the results for the closest single-mass isotope to $^{101}Sn$ with the experimental levels of $1/2^{+}$, $3/2^{+}$ and $11/2^{-}$ in the literature. ANN estimates for this $^{109}Sn$ isotope are 0.546, 0.518 and 0.553 keV for set-1, set-2 and set-3 for the $1/2^{+}$ level, respectively. The corresponding experimental value is 0.545 keV. For another level, $3/2^{+}$ level, the values obtained with ANN are 0.673, 0644 and 0.698 keV, respectively, and the experimental value is 0.664 keV. The final value for the estimation of $^{109}Sn$ by ANN is the energy of the $h_{11/2}$ level, which are 1.221, 1.300 and 1.259 keV, respectively. The experimental value of this level in the literature is 1.269 keV. As can be again seen that ANN is a suitable tool for obtaining single particle energies.

Although the ANN results of all three sets are acceptable, it is seen in Table 2 that different neutron spe values are obtained in the calculations made with different sets. While the ANN results of the 101Sn isotope of those numbered 2 and 3 of these sets are close to each other, the results of set number 1 are different from the other two. It is even seen that the energy of the $1/2^{+}$ level in this set is even below the $11/2^{-}$ level. However, it will be seen in the next section that the results of the calculations made with these separate sets give better results compared to the SM calculations performed using the current neutron spe values in the literature.

\subsection{Shell Model Calculations} 
Using the neutron spe values we obtained by ANN method, we calculated the low-lying energy, spin and parity values of the ground and excited states for all even and odd Sn isotopes in the mass range of 102-108 through the nuclear shell model calculations. 

\begin{figure}
\includegraphics[width=18cm]{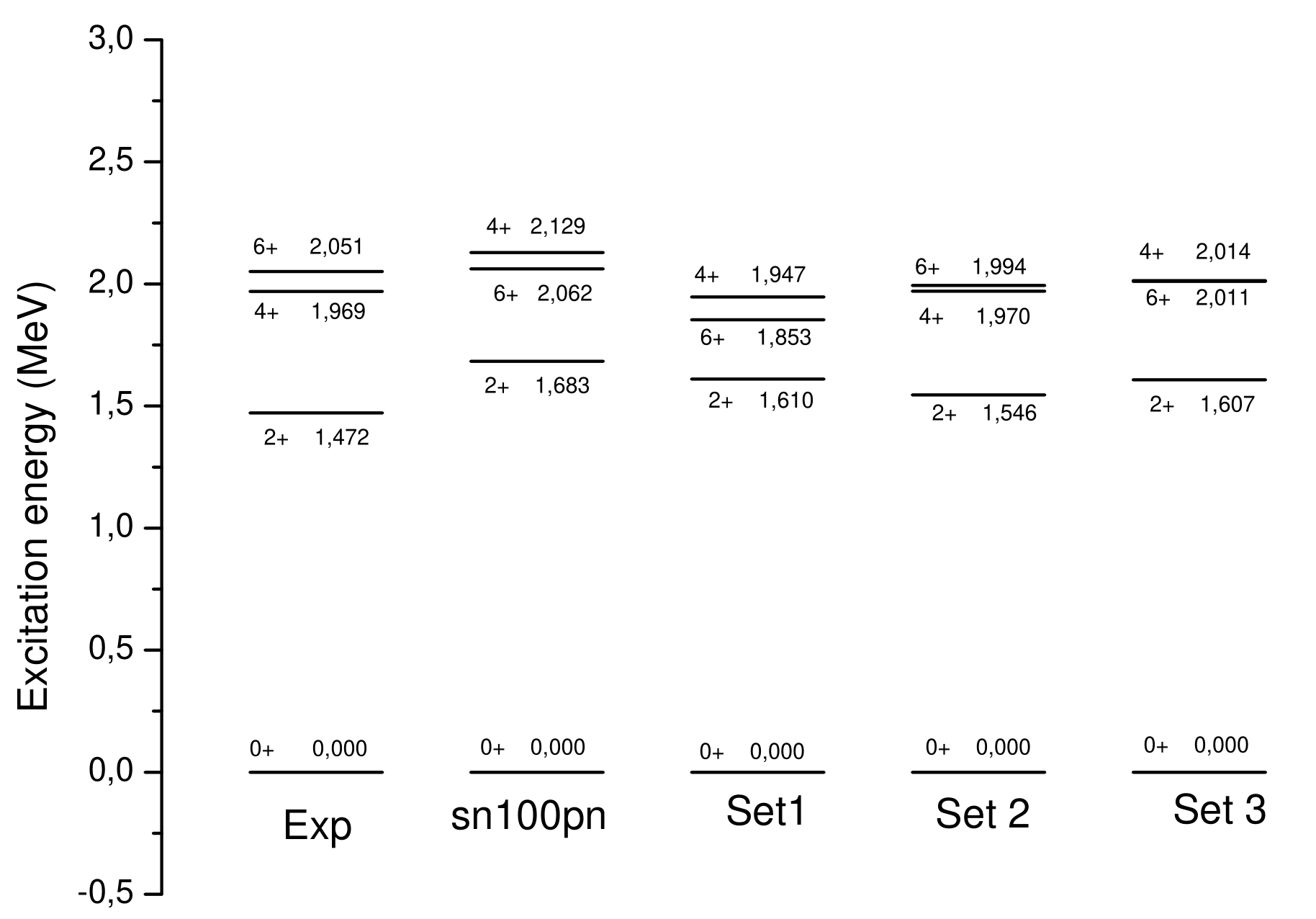}
\caption{\label{Fig4} Experimental and calculated low-lying energy states of $^{102}Sn$ isotopes}
\end{figure}

The results of the calculations we made for $^{102}Sn$, of which very few experimental data are available in the literature are given in Fig.\ref{Fig4}. As can be clearly seen from the figure that the ground-state spin and paritiy can be obtained correctly in all four theoretical calculations. It was seen that the values of the first $2^{+}$ state in the shell model calculations we performed using three different neutron spe values obtained with ANN (set-1, set-2, set3) were better than those of the current sn100pn in the literature. Among them, the closest to the experimental value is the neutron spe values belonging to the set-2. When we look at the energy of the first $4^{+}$ excited state, it is seen that the results of the calculations performed with the neutron spe values obtained by ANN are closer to the experimental results compared to the sn100pn results. The closest of these is the one that belongs to the set-2 again. Finally, when we examine the calculated energies of the first $6^{+}$ state, it was seen that the sn100pn results were better than the values of the set-1, set-2 and set-3. However, the results of the set-2 and set-3, although not better than sn100pn, are close to this value. It was seen that the ordering of the states was obtained correctly with the set-2. It can be concluded that the results of shell model calculations using neutron spe values obtained by ANN are generally better than the results obtained with sn100pn and closer to the experimental values.

\begin{figure}
\includegraphics[width=18cm]{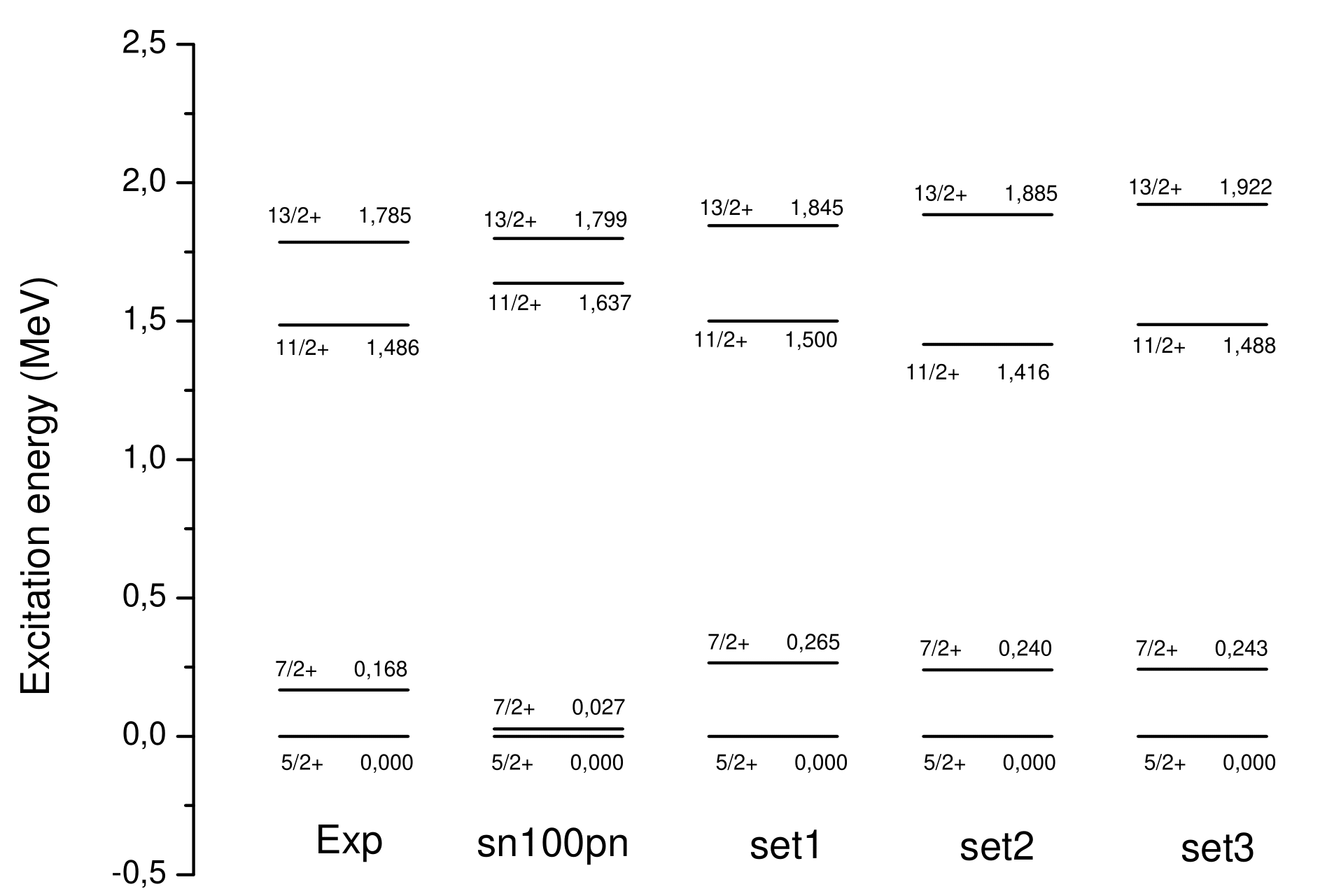}
\caption{\label{Fig5} Experimental and calculated low-lying energy states of $^{103}Sn$ isotopes}
\end{figure}

The results of the calculations we made for $^{103}Sn$, of which only a few of the experimental data are available in the literature, are given in Fig.\ref{Fig5}. The ground-state spin and parity can be obtained correctly in all four calculations as $5/2^{+}$. The values of the first $7/2^{+}$ state in the shell model calculations we performed using three different neutron spe values obtained with ANN (set-1, set-2, set3) were better than those of the current sn100pn in the literature. Among them, the closest to the experimental value belongs to the set-2. When we look at the energy of the first $11/2^{+}$ excited state, it is seen that the results of the calculations performed with the neutron spe values obtained by ANN are closer to the experimental results compared to the sn100pn results. The closest of these is the one that belongs to the set-3. Finally, when we compare the calculated energies of the $13/2^{+}$ level, it is seen that the result obtained from the use of sn100pn is better. From this, it can be concluded that the results of shell model calculations using neutron spe values obtained by ANN are generally better than the results obtained with sn100pn and closer to the experimental values. 

\begin{figure}
\includegraphics[width=18cm]{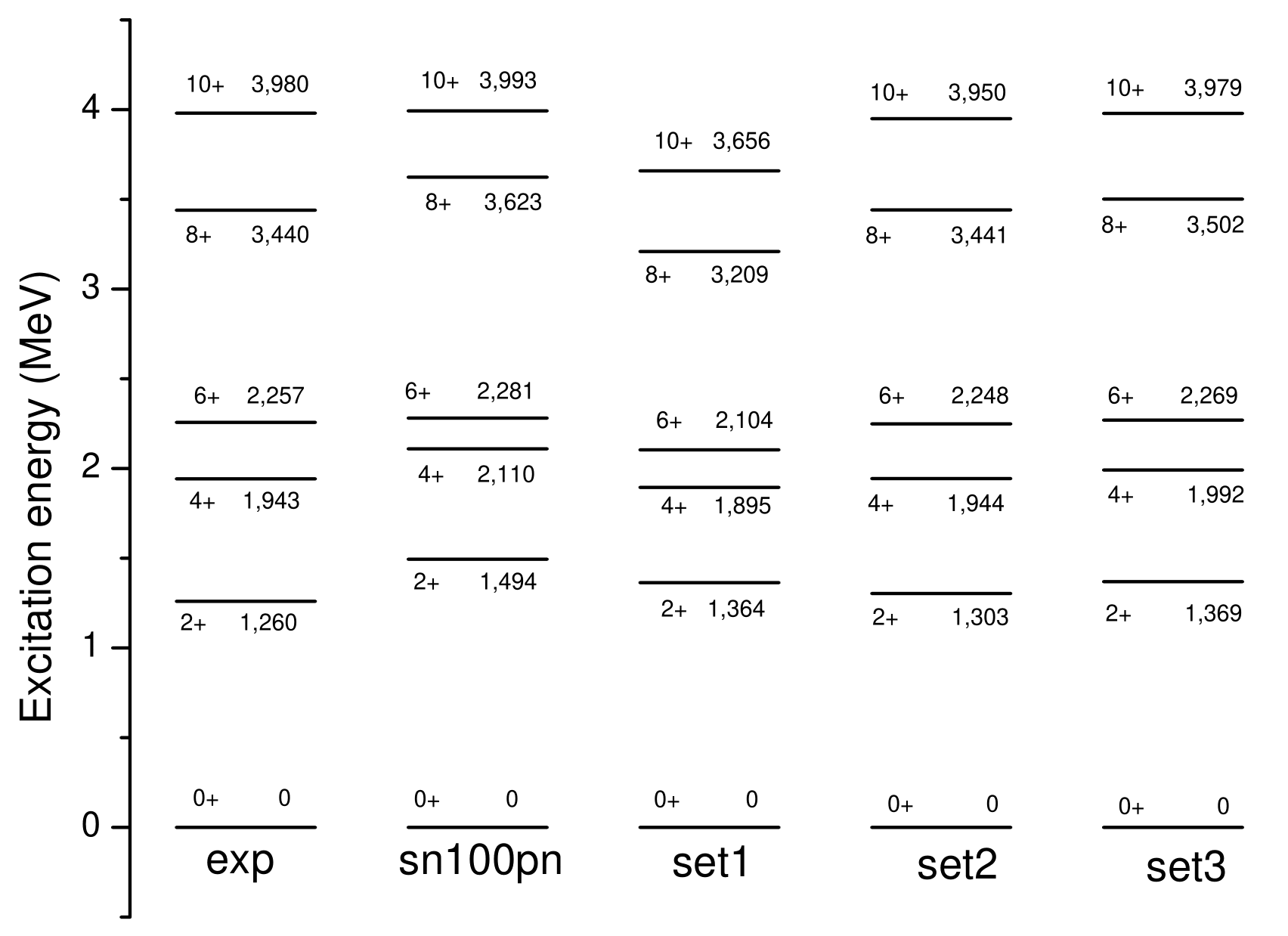}
\caption{\label{Fig6} Experimental and calculated low-lying energy states of $^{104}Sn$ isotopes}
\end{figure}

In the shell model calculations performed for the $^{104}Sn$ isotope, it is seen that the results obtained from the different sets produced with ANN generally give the energies of the first states better (Fig.\ref{Fig6}). As can be seen in the table that the ground-state spin and parity can be obtained correctly in all four calculations. The energies of the first excited $2^{+}$ and $4^{+}$ states from the set-1, set-2 and set-3 are better than sn100pn. Among these, the closest experimental result was obtained from the calculations made with the set-2. Also, the energies of the first excited $6^{+}$ and $8^{+}$ levels could be obtained closer to the experimental value by using the set-2 and set-3. The energy of the first excited $10^{+}$ level is also calculated closer to the experimental value by set-3. As a result, it is seen that the set-3 generally gives more consistent results with the experimental values in the calculations of the low-lying excited states of the $^{104}Sn$ isotope.

\begin{figure}
\includegraphics[width=18cm]{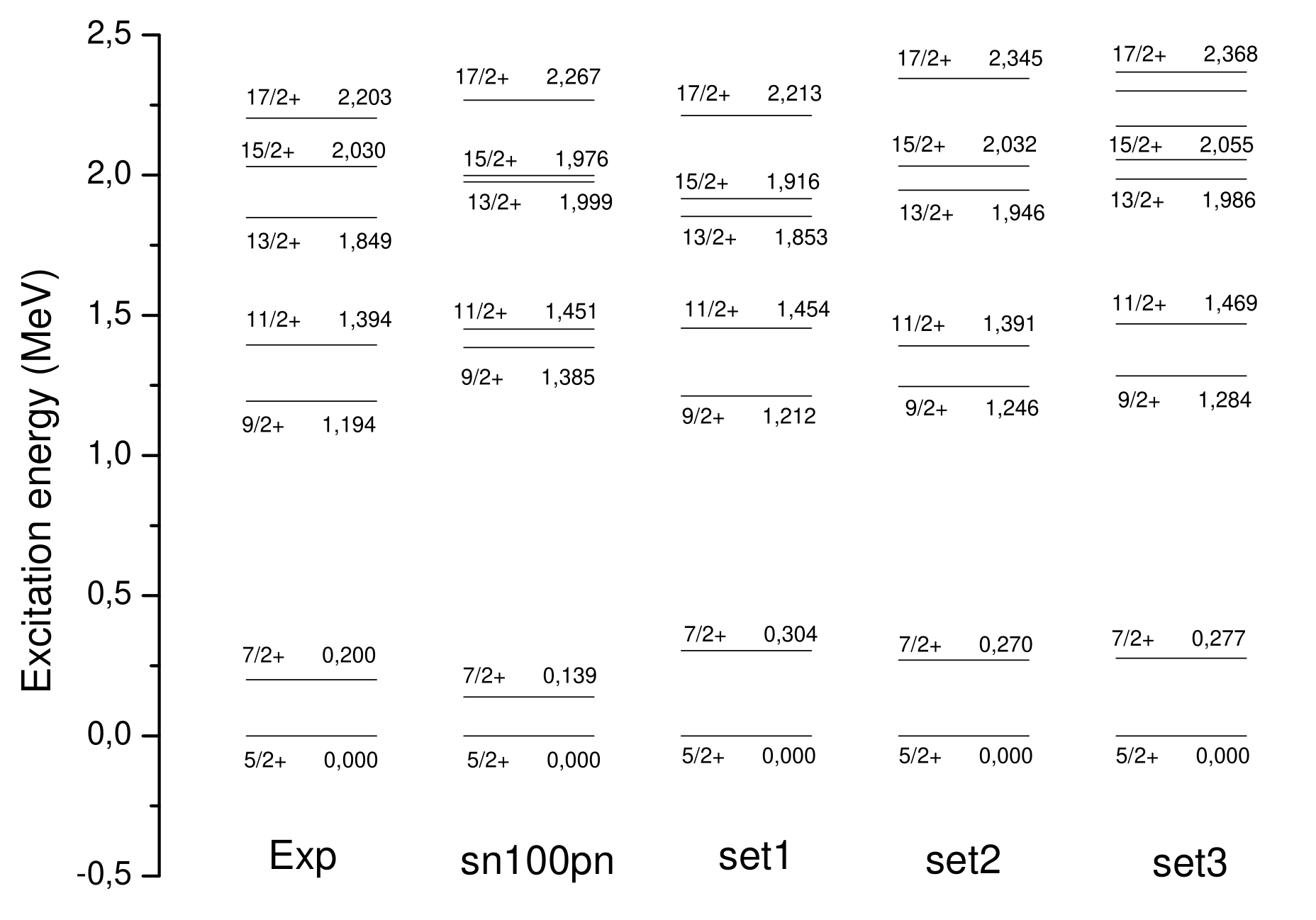}
\caption{\label{Fig7} Experimental and calculated low-lying energy states of $^{105}Sn$ isotopes}
\end{figure}

The results of our calculations for the $^{105}Sn$ isotope are shown in Fig.\ref{Fig7}. The ground state spin and parity of this isotope were correctly calculated as $5/2^{+}$ in all theoretical calculations. For the first $7/2^{+}$ level, sn100pn result is slightly better than those of different sets from ANN. The energy of the first excited $9/2^{+}$ state was obtained with 3 different sets produced with ANN, giving results closer to experimental values. The set-1 gives the best result for this state. The $11/2^{+}$ energy is better predicted by set-3. For the energies of the first and second $13/2^{+}$ states, results closer to the experimental values were obtained in the set-1 and set-2 calculations compared to the sn100pn calculations. Also for the first $13/2^{+}$, set-3 gives closer results than sn100pn. Among these 3 sets, it is seen that the results given by set-1 are closer to the experimental values. The set-2 and set-3 gave better results for the first excited 15/2+ state, while the result of set-1 was found to be better than the others for the first excited $17/2^{+}$ state.

\begin{figure}
\includegraphics[width=18cm]{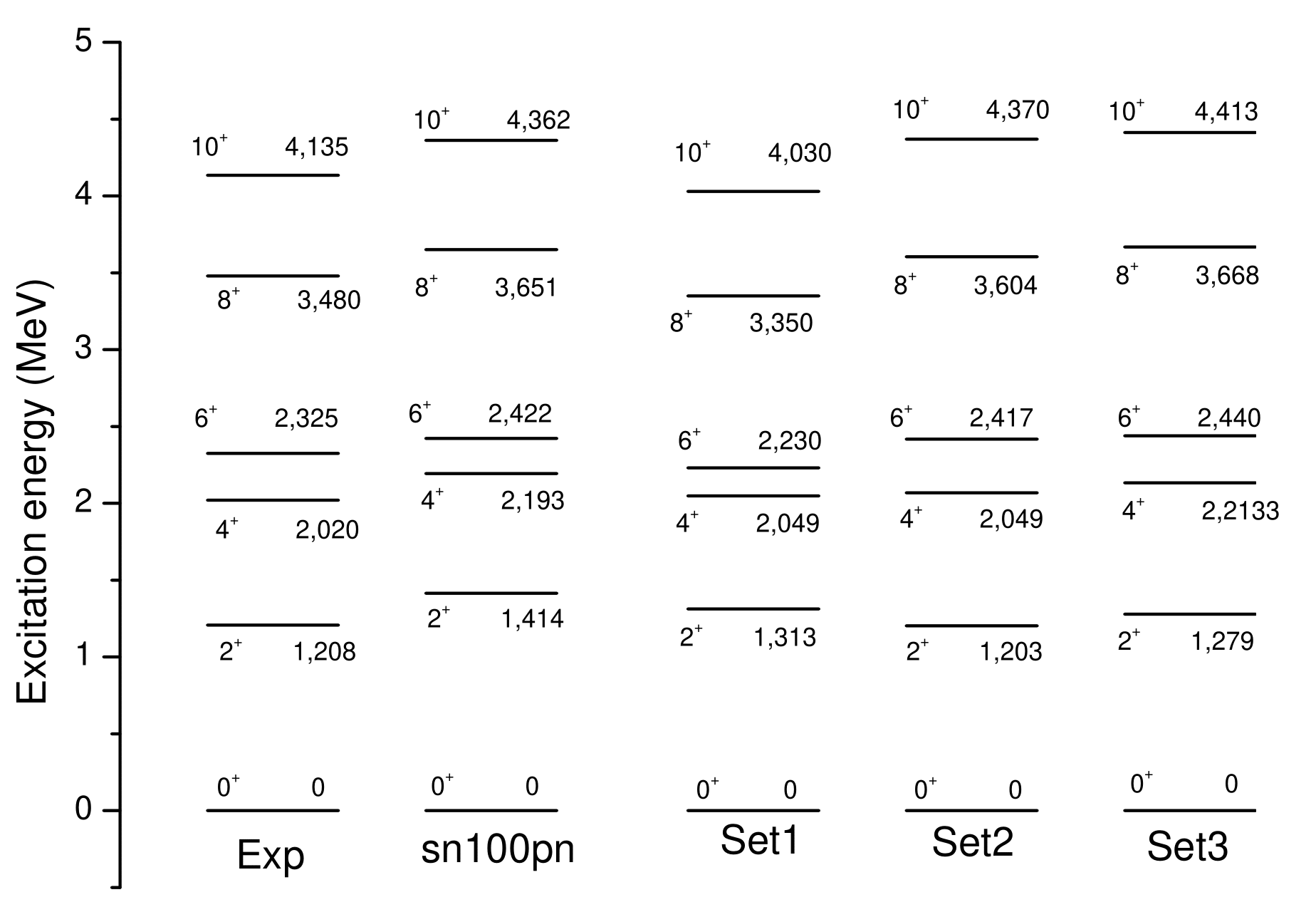}
\caption{\label{Fig8} Experimental and calculated low-lying energy states of $^{106}Sn$ isotopes}
\end{figure}

The results of the shell model calculations for the $^{106}Sn$ isotope are shown in Fig.\ref{Fig8}. Ground-state spin and parity have been obtained correctly in all theoretical calculations. The energies of the first excited $2^{+}$ and $4^{+}$ states were obtained with sets produced with ANN better than sn100pn. Of these, set-2 and set-1 gave results closer to experimental values than others for $2^{+}$ and $4^{+}$, respectively. Although the energy of the first excited 6+ state is better calculated by the set-1 and set-2, the result of the set-3 is also very close to this value. In the calculations performed by the set-1 and set-2, the energy values of the first excited $8^{+}$ state were calculated closer to the experimental value compared to the others. Finally, in calculating the energy of the first excited $10^{+}$ state, the set-1 is the most successful.

\begin{figure}
\includegraphics[width=18cm]{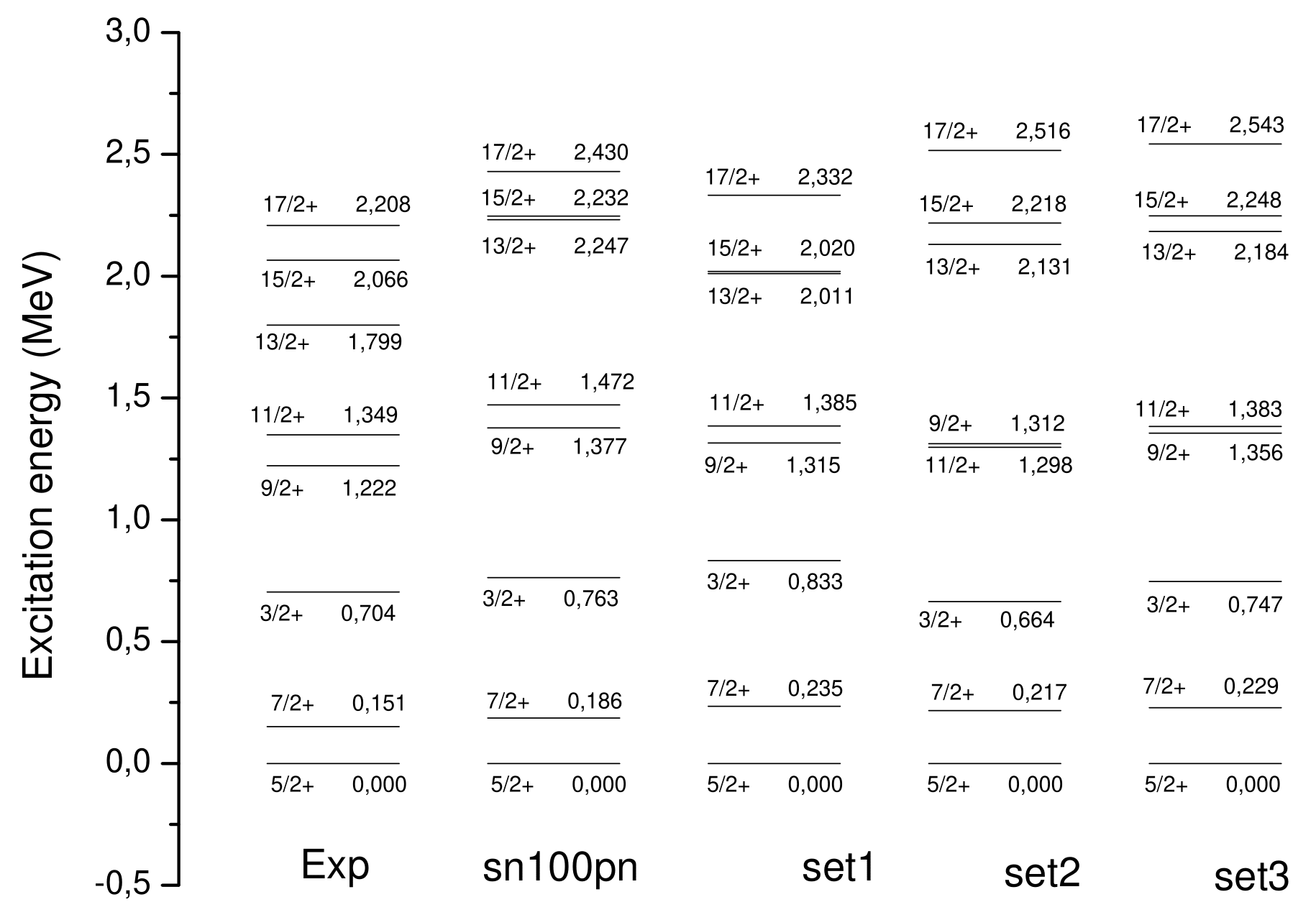}
\caption{\label{Fig9} Experimental and calculated low-lying energy states of $^{107}Sn$ isotopes}
\end{figure}

As a result of examining the results of the shell model calculations performed for the $^{107}Sn$ isotope is given in Fig.\ref{Fig9} with the experimental results. It is seen that the calculations of three different sets obtained with ANN are generally better than the results of sn100pn calculation. If we look at the comparison of each of these sets with sn100pn individually, we find that the result for the first excited $7/2^{+}$ and $3/2^{+}$ energies from the set-1 is worse than sn100pn, while for all other excited levels, the set-1 gives better results. When we make the similar comparison for set-2, it appears that sn100pn gives better results for the first excited $7/2^{+}$, $3/2^{+}$ and $17/2^{+}$ states.

The results of the shell model calculations for the $^{108}Sn$ isotope are shown in Fig.\ref{Fig10}. The ground state spin and parity values were obtained in harmony with the experimental values in all theoretical calculations. The energy of the first excited $2^{+}$ state was calculated using the set-1 at almost the same value as sn100pn. These obtained values are closer to the experimental values compared to the other results. For the first excited $4^{+}$ state, the results of the sn100pn calculations are better, but the results from the set-1 are the best among the other sets and close to the experimental value. The energy of the first excited $6^{+}$ state was obtained with the set-3 closest to the experimental data. The set-2 results are also comparable to sn100pn results for this level. Finally, according to the calculations with the set-2 and set-3, the energy of the first excited $5^{+}$ state gave better results than sn100pn.

\section{Conclusion}
In this study, we obtained the low-lying energy states of the $^{101}Sn$ isotope with the ANN method and obtained the neutron single-particle energies for the shell model calculations to be carried out in this region. For this purpose, we conducted ANN training with the energy level values of isotopes with the available experimental spectra in the $^{100}Sn$ region and we obtained the energies of $5/2^{+}$, $7/2^{+}$, $1/2^{+}$, $3/2^{+}$ and $11/2^{-}$ states for $^{101}Sn$. Using these values we obtained and the experimentally known spectrum of 101Sn, we compiled neutron single-particle energies. These neutron single-particle energies are used in the modification of the sn100pn interaction, which is widely used in the shell model calculations in the $^{100}Sn$ region in the literature. We calculated the ground state spin/parity values and low-level excited states of the $^{102-108}Sn$ isotopes with the calculations we performed within the scope of the nuclear shell model. The results of the original sn100pn were compared with the results of the modified sn100pn. We found that the results from the modified sn100pn yielded results that were more in line with the experimental spectrum overall. If we examine the results of the calculations performed with different sets for $^{102-108}Sn$ isotopes, we can make a general comparison. We can do this comparison by looking at the number of excited states that are closer to the experimental data. Accordingly, we see that the results obtained from set-2 for $^{102}Sn$ are more in agreement with the experimental results. For $^{103}Sn$, set-3 results are more preferable. The results of the calculations performed with set-2 for $^{104}Sn$ are in full agreement with the experimental data at all excited energy states. While set-1 and set-2 may be preferred for $^{105}Sn$, it is seen that set-2 may be better for $^{106}Sn$. For the $^{107}Sn$ isotope, although set-1 was better, it caused degeneracy at $15/2^{+}$ and $13/2^{+}$ levels. Finally, when the $^{108}Sn$ isotope is examined, it is seen that set-3 is better than the other sets. After these reviews and seeing that sets with ANN generally give better results than sn100pn, it is concluded that ANN-supported single-particle energies can be possibly obtained confidently.

\section*{Data availability statement}
The data that support the findings of this study are available from the corresponding author upon reasonable request.

\newpage

\begin{figure}
\includegraphics[width=18cm]{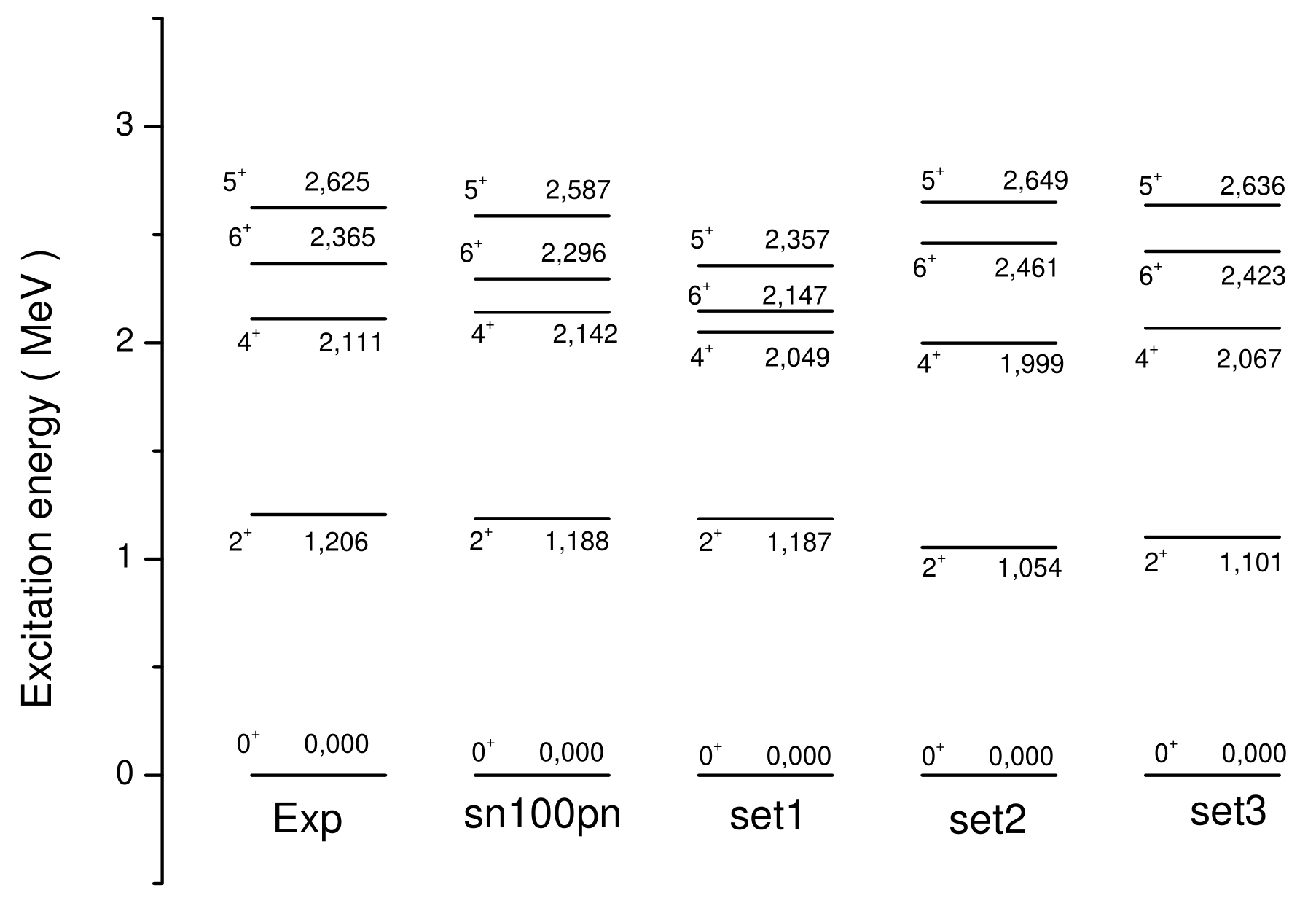}
\caption{\label{Fig12} Experimental and calculated low-lying energy states of $^{108}Sn$ isotopes}
\end{figure}

\end{document}